\documentclass[10pt, conference]{IEEEtran}

\usepackage{cite}
\usepackage{amsmath,amssymb,amsfonts}
\usepackage{algorithmic}
\usepackage{balance}
\usepackage{enumitem}
\usepackage{framed}
\usepackage{graphicx}
\usepackage[colorlinks=true, allcolors=blue]{hyperref}
\usepackage{textcomp}
\usepackage{xcolor}
\def\BibTeX{{\rm B\kern-.05em{\sc i\kern-.025em b}\kern-.08em
    T\kern-.1667em\lower.7ex\hbox{E}\kern-.125emX}}

\begin{document}

\title{Uncovering Code Insights: Leveraging GitHub Artifacts for Deeper Code Understanding}
\author{
\IEEEauthorblockN{Ziv Nevo, Orna Raz, Karen Yorav}
\IEEEauthorblockA{\textit{IBM Research -- Israel} \\
\{nevo, ornar, yorav\}@il.ibm.com}
}

\maketitle

\begin{abstract}
Understanding the purpose of source code is a critical task in software maintenance, onboarding, and modernization. While large language models (LLMs) have shown promise in generating code explanations, they often lack grounding in the broader software engineering context. We propose a novel approach that leverages natural language artifacts from GitHub---such as pull request descriptions, issue descriptions and discussions, and commit messages---to enhance LLM-based code understanding. Our system consists of three components: one that extracts and structures relevant GitHub context, another that uses this context to generate high-level explanations of the code's purpose, and a third that validates the explanation. We implemented this as a standalone tool, as well as a server within the Model Context Protocol (MCP), enabling integration with other AI-assisted development tools. Our main use case is that of enhancing a standard LLM-based code explanation with code insights that our system generates.  To evaluate explanations' quality, we conducted a small scale user study, with developers of several open projects, as well as developers of proprietary projects. Our user study indicates that when insights are generated they often are helpful and non trivial, and are free from hallucinations.  
\end{abstract}


\section{Introduction}
\label{intro}
Modern software systems are increasingly complex, and understanding the purpose of existing code is a major challenge in software maintenance, on-boarding, and modernization. Traditional code explanation tools focus on what the code does—--its execution semantics—--but often fail to explain \emph{why} the code exists in the context of the application’s features, architecture, or evolution.

Recent advances in LLMs have enabled impressive capabilities in code generation and explanation. However, these models typically operate in isolation from the rich natural language (NL) context available in software repositories. GitHub, for example, contains a wealth of NL artifacts such as pull request (PR) descriptions, issue descriptions and discussions and commit messages, that capture the rationale behind code changes, implicit technical debt, and evolving requirements. 

In this work, we develop a system that leverages these artifacts to enhance code understanding. Our approach consists of three components: (1) a context extractor that retrieves and structures relevant GitHub artifacts using the GraphQL API, (2) an LLM-based explanation generator that uses this context to produce high-level, purpose-driven explanations of code, and (3) a validator to asses the quality of the produced explanation. The third component includes an LLM-as-a-Judge (LaaJ) validator that we developed in order to filter out insights that may be of low quality (for example because they contain hallucinations).  We implemented these tools in an MCP server, enabling seamless use in broader AI-assisted workflows.

\begin{figure}[ht]
    \centering
    \includegraphics[width=1\linewidth]{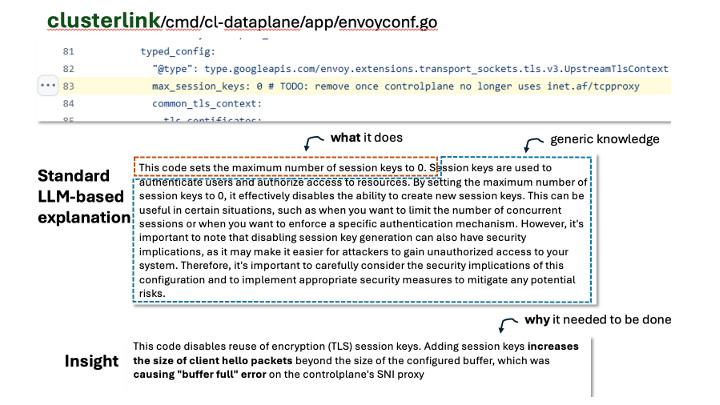}
    \caption{Example of a standard explanation of the highlighted code snippet, along with our system's insight.}
    \label{fig:insight-example}
\end{figure}

Figure \ref{fig:insight-example} provides an example of a standard explanation along with an insight generated by our system. The standard explanation details the code functionality along with general information about that functionality, in this case that the code sets the number of session keys to 0 along with an explanation of what session keys are and what it means to set their number to 0. The code insight, however, provides the rationale for why this functionality is there in the first place. In our example, it was necessary to disable session keys because of a problem that their addition was causing. 

We found it important to generate high quality insights or generate no insights at all. For this purpose our insight generator includes an LLM-as-a-Judge (LaaJ) filtering component. To develop this LaaJ we experimented with various potential implementations, using structured prompts to assess explanation quality along two dimensions: well-formedness and groundedness. Our experiments show that a two-step evaluation process that first extracts claims and then verifies them yields more reliable assessments.

We demonstrate the effectiveness of our system on the use case of code explanation, by providing evidence of improving explanations. Our technology is especially effective in the domain of software systems maintenance. Understanding why a particular code is implemented the way it is, as well as being aware of past defects that the implementation avoids, is essential to many challenging code maintenance scenarios. For example, identifying code failures and avoiding regression failures. In addition, we find that the GitHub context often gives insight into how certain components contribute to the overall application requirements.

Evaluating our system is non-trivial due to the absence of ground truth code insights. Human-written references are often inconsistent or include information not present in the code or its context. We conducted a user study with developers of multiple open source and proprietary projects. The results of the study are encouraging. Users consistently indicate that the insights are often a helpful addition to the standard explanation. When insights are not available, the standard explanation is unharmed. None of the users found any hallucinations in the insights. 

The contributions of this paper are a novel system for enhancing code understanding with grounded insights, a context extractor, and a LaaJ for filtering unreliable insights. 

The rest of the paper is structured as follows: Section~\ref{related} discusses related works, Section~\ref{solution} provides a high-level description of our solution, Section~\ref{implementation} gives details on how we implemented our ideas, Section~\ref{evaluation} summarizes our user-study evaluation, while Section~\ref{conclusions} concludes this work.

\section{Related Work}
\label{related}
\subsection{AI for Code Understanding}

LLMs have demonstrated strong performance in code related tasks, such as code generation and code explanation. Tools like GitHub Copilot \cite{microsoft2025copilot} and Amazon CodeWhisperer \cite{amazon2025codewhisperer} use transformer-based models to suggest code completions and explanations within IDEs, significantly improving developer productivity \cite{10.1145/3633453}. However, these tools primarily focus on syntactic and semantic code patterns, often lacking awareness of the broader software engineering context.

Several studies have evaluated the capabilities and limitations of LLMs in code-related tasks. For example, Tang et al. \cite{10507163} analyzed ChatGPT’s performance on LeetCode problems and found that while it performs well on familiar tasks, it struggles with novel or complex problems due to limited contextual understanding. This highlights the need for grounding LLMs in external, task-specific context to improve reliability and reduce hallucinations. The context creation tool of our system is an example. 

\subsection{Contextual Code Explanation}

Prior work has explored the use of version control and issue tracking systems to enhance code comprehension. For instance, commit messages and issue discussions have been used to trace the rationale behind code changes, for example in Claude Code \cite{claudeCodeGuide2025} Q\&A (question answering). However, few systems integrate this information into LLM-based explanation pipelines. Our approach builds on this idea by systematically extracting and structuring GitHub artifacts to serve as input context for LLMs.

There are many ways to enrich the code itself. Some are based on various kinds of static and dynamic analysis, for example call graph data \cite{makharev2025codesummarizationfunctionlevel}. Others may be based on documentation and user manuals. Our work uses github artifacts.

\subsection{Online evaluation of AI-Generated Explanations}

Evaluating AI-generated code explanations remains a challenge. Offline evaluation methods often fail to capture the diversity of valid explanations, especially when multiple interpretations are possible \cite{10.1145/3633453}. Recent work has proposed using LLMs themselves as evaluators, but naive prompting can lead to unreliable judgments. Our two-step LaaJ method improves evaluation fidelity by externalizing the model’s reasoning process. Thus, it enables to filter insights so users receive insights that have increased probability to be useful and helpful. 



\section{GitHub-powered Code Understanding}
\label{solution}
Modern software repositories, particularly those hosted on platforms like GitHub, contain a wealth of natural language (NL) artifacts that go far beyond the source code itself. These include pull request (PR) descriptions, issue descriptions and discussions, commit messages, Wiki pages, README files, GitHub Discussion pages and project documentation. Such artifacts often capture critical software engineering (SWE) context, such as architectural decisions, implementation decisions, root cause of bugs, technical debt, feature requirements, and user experience considerations. Leveraging this rich textual ecosystem can significantly enhance the ability of AI systems to explain not just what a piece of code does, but why it exists, why it is written the way it is, and how it fits into the broader application context.

\begin{figure}[ht]
    \centering
    \includegraphics[width=1.0\linewidth]{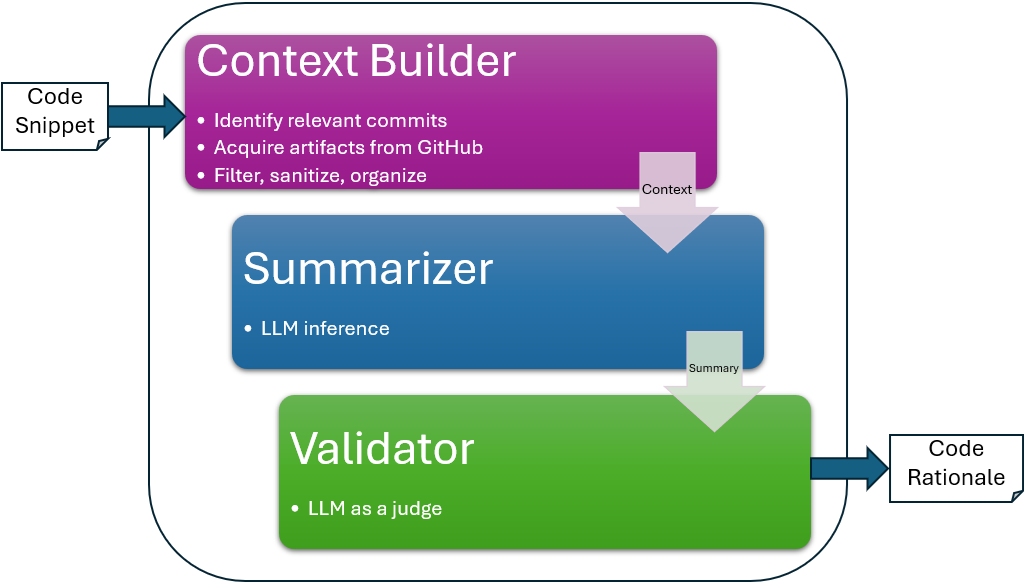}
    \caption{High-level architecture of proposed solution.}
    \label{fig:high-level-arch}
\end{figure}

We propose a three-stage approach that harnesses GitHub's NL artifacts to improve code understanding, as depicted in Figure~\ref{fig:high-level-arch}. 

The first component in our system, the \emph{Context Builder},  extracts texts from GitHub repository artifacts that are relevant to the given piece of code. It then organizes them in a hierarchical structure, maintaining the relationships between the artifacts. This structure is the input to the second component. In a sense, the Context Builder mimics the manual work of developers navigating unfamiliar or legacy codebases.

The second component, the \emph{Summarizer}, uses an LLM to generate a high-level explanation of the code, utilizing the context extracted in the first stage. Unlike traditional code summarization tools that focus on execution semantics, this system aims to explain the \emph{purpose} of the code in the context of the application's architecture, features, and evolution.

The LLM is prompted with both the code and the structured context, enabling it to answer questions such as:
\begin{itemize}
    \item What feature or requirement motivated this code?
    \item What bugs or technical debt does it address?
    \item How has the implementation evolved over time?
\end{itemize}

The LLM response is now given to the \emph{Validator} (based on yet another LLM), which checks the high-level code explanation for well-formedness and the lack of hallucinations. A well-formed, hallucination-free summary is then given to the user.

This approach supports several practical scenarios:
\begin{itemize}
    \item \textbf{Understanding unfamiliar or legacy code:} Developers can quickly grasp the rationale behind complex or aged code without manually sifting through GitHub history.
    \item \textbf{Onboarding new team members:} The explanation tool acts as a virtual mentor, providing contextualized insights akin to those from experienced colleagues.
    \item \textbf{Catching-up with recent code changes:} When a developer returns to a piece of code they haven't touched for a while, a summary of all recent changes can be provided.
    \item \textbf{Preventing regression errors:} When modifying code, both human developers and AI agents benefit from understanding the historical and architectural motivations behind existing implementations—especially when the code is non-intuitive due to accumulated patches or evolving requirements.
\end{itemize}

By grounding code explanations in authentic project artifacts, our approach bridges the gap between low-level code semantics and high-level software engineering intent. This not only enhances developer productivity but also aligns with the broader goals of AI-assisted software modernization.

\section{Implementation Details}
\label{implementation}
Recall that our system is made of three components: the Context Builder,  the Summarizer, and the Validator. In this section we provide more details on the implementation of each component, with a focus on scalability, modularity and fidelity.

\subsection{Context Builder}

The first component in our system extracts, filters and organizes relevant contextual information from GitHub repositories. Given a code snippet, it sequentially performs the following steps:

\begin{enumerate}
    \item Uses \texttt{git log} to trace the commit history associated with the snippet.
    \item Filters out trivial commits.
    \item For each remaining commit, identifies the corresponding PR and any linked issues.
    \item Organizes the extracted information in a hierarchical data structure, preserving relationships between commits, PRs, and issues.
    \item Extracts NL descriptions from these commits, PRs and issues, applying filters to remove malformed or excessively verbose content.
    \item Builds an LLM-ready structured context from the hierarchical data structure and the filtered texts.
\end{enumerate}

More details on each step are given below.
\subsubsection{Extracting commit history}
Given a code snippet, that is, a file in a git repository clone and line numbers marking the snippet beginning and its end, we run \texttt{git log} and use its \texttt{-L} flag~\cite{gitlog} to specify the given file and the code block to examine. The output of this command is a list of commits which modified the snippet, including their SHA signature, the associated user comment, and a list of relevant diff hunks. We parse this output and store these data items for each commit.
\subsubsection{Filtering-out trivial commits}
Given a list of commits, we analyze each of the associated diff hunks and decide if it is trivial or not. A commit is trivial if \emph{all} its code-snippet-modifying diff hunks are trivial. A diff hunk is trivial if it only deletes lines, if it only modifies code comments, if it only modifies text in strings or if this is a simple variable renaming. Trivial commits are removed from the list provided to the next steps.
\subsubsection{Extracting related PRs and Issues}
We use GitHub GraphQL API~\cite{github-graphql}, and build a GraphQL query optimized for large repositories that may contain extensive commit histories and numerous linked artifacts. The query is designed to:

\begin{itemize}
    \item Traverse from commits to associated pull requests (PRs), extracting PR number, title, body, and URL.
    \item Follow links from PRs to closing issues and timeline events, including cross-referenced and connected issues.
    \item Extract issue metadata such as number, title, body, and URL.
\end{itemize}
The query is designed to fetch information on multiple commits in a single call, minimizing network traffic, enhancing performance, and avoiding rate limits. Pagination parameters (e.g., \texttt{first: 100}) are tuned to balance completeness and performance.
\subsubsection{Organizing data hierarchically}
Artifacts extracted by the GraphQL query are organized in a hierarchical data structure, where for each extracted PR, both its commits and its linked issues are listed as children. If a commit has no originating PR, it is listed separately. Each artifact is listed with its unique identifier (number for PRs and issues, SHA for commits), its URL and its title and body texts. This preserves provenance and later enables the LLM to reference all artifacts accurately.

\subsubsection{Extracting relevant texts}
This step ensures that texts in the hierarchical data structure are both relevant and efficient for downstream LLM processing. We apply several filtering and formatting strategies:

\begin{itemize}
    \item \textbf{Well-formedness filtering:} We discard PRs and issues with empty or malformed bodies (e.g., containing nothing but emojis).
    \item \textbf{Length truncation:} Descriptions exceeding a configurable token limit are truncated to avoid overwhelming the LLM context window.
    \item \textbf{Template-aware summarization:} For repositories using structured PR or issue templates, we extract only the most informative sections (e.g., summary, motivation).
    \item \textbf{Noise reduction:} Boilerplate text, checklists, and unrelated discussions are removed using regular expressions and heuristic rules.
\end{itemize}

\subsubsection{Building an LLM-ready structured context}
The last step builds a single text string with all relevant information to serve as the context for the LLM. Here we format each item in the hierarchical data structure, using hypertext tags, titles and indentation, to reflect the relationships between PRs, commits and issues. This (now serialized) structure aids downstream LLMs in referencing artifacts accurately (e.g., by PR or issue number).

Together, all these steps are critical to minimizing token usage, reducing distraction, and guiding the LLM towards a high-signal content.

\subsection{Summarizer}

The second component in our system puts the context from the Context Builder into a use-case specific LLM prompt, and sends it to any given LLM.

Figure~\ref{fig:exmaple-prompt} shows an example of such a prompt for the code-understanding use-case. Note how the context information includes both the code snippet and the relevant artifacts retrieved from GitHub. Also note how PR context includes both related issues and related commits. Finally, note how \emph{begin} and \emph{end} tags are used to organize context in a PR-centric way.

\begin{figure}[ht]
\begin{framed}
\scriptsize
\begin{verbatim}
You are a software developer with experience
working on large projects and you are helping a
junior developer understand a code snippet.
You will be given a code snippet and you need
to explain why this code is needed where it is,
and what it is for. To aid you, you will also
be given a list of pull requests (PRs) that are
in the log for this code snippet, appearing in
chronological order. Your answer should be a
short and to the point explanation of the
purpose of the code snippet. Before you answer,
consider what is important for the junior
developer to know and understand. You may
mention PR and issue numbers in your response,
but only the most important ones.

[begin context information]
[begin code snippet]
func newSGSplitSubnet(name string,
    configs map[string]*vpcmodel.VPCConfig
) linter {
    return &filterLinter{
        basicLinter: basicLinter{
            configs:     configs,
            name:        name,
            enable:      false,
        },
        layer:   vpcmodel.SecurityGroupLayer,
        checkForFilter: findRedundantRules
    }
}
[end code snippet]
\end{verbatim}
{\color{blue}\begin{verbatim}
[begin Pull Request #742]
Lint syntactically redundant rule. 
Added a lint for syntactically redundant rules.
A rule is syntactically redundant in `SG` if
other rules in the table imply it.

Issues relevant to this PR:
Issue #115: Add a warning for useless sg rules
based on the VPC APs

Commits relevant to this PR:
commit #1: lint syntactic redundant rule (#742)
added lint for syntactically-implied SG rules
[end Pull Request #742]
\end{verbatim}}
\begin{verbatim}
[end context information]

Answer:
\end{verbatim}
\end{framed}
\caption{An example prompt, containing GitHub context (blue).}
\label{fig:exmaple-prompt}
\end{figure}

It is the job of the Summarizer to use the right prompt for a given use-case, whether this is code understanding, summarizing code history, identifying possible pitfalls or enriching RAG indexing.

In order to keep explanations concise, the number of output tokens is limited to several hundreds. If the context appears to contain many PRs and commits, more output tokens are allowed, however the addition of tokens per change decline as the number of changes grow (using a square root function).

The code calls the LLM to generate the summary by using the API of \emph{watsonx.ai as a Service}~\cite{watsonx-ai-api}. Watsonx.ai hosts many open-source models, which eases making comparisons and deciding on the best model for a given use-case.

\subsection{Validator}

To ensure the reliability of explanations generated by our system, we developed a validation component that operates at runtime to assess whether an LLM-generated explanation is suitable for presentation to the user. This component adopts the paradigm of \emph{LLM-as-a-Judge}, wherein a language model is tasked with evaluating the quality of another model's output.

Although our system provides the judge with contextual information extracted from GitHub, the evaluation methodology is generalizable and can be applied to explanations generated without such context. The judge assesses two key dimensions: (i) whether the explanation is well-formed, and (ii) whether it contains hallucinated claims, i.e. statements not supported by the provided context.

We defined a four-point scoring rubric to guide evaluation:

\begin{itemize}
    \item \textbf{0:} Explanation is acceptable.
    \item \textbf{1:} Contains a single hallucinated claim (minor factual error).
    \item \textbf{2:} Contains multiple hallucinated claims.
    \item \textbf{3:} Malformed (e.g., repetitive, off-topic, or merely restates the context).
\end{itemize}

Initial experiments with a naive judge, prompting the LLM to directly assess the explanation against the context, yielded inconsistent results. We attribute this to the model attempting to reason silently, without externalizing its thought process. To address this, we adopted a structured evaluation strategy: the LLM is first asked to enumerate the factual claims made in the explanation, and then assess each claim for groundedness in the context (including both the code snippet and related GitHub artifacts). This approach significantly improved judgment reliability by encouraging explicit reasoning.

We implemented and evaluated four variants of the judge:

\begin{enumerate}[label=\textbf{Judge\arabic*:}, leftmargin=*, itemsep=0.5em]
    \item A single prompt instructing the LLM to score the explanation using the defined rubric.
    \item A single prompt that instructs the LLM to execute a sequence of tasks: (i) assess well-formedness, (ii) list claims, (iii) evaluate each claim for hallucination, and (iv) output the score and hallucinated claims.
    \item A single prompt with two distinct tasks: one for well-formedness (pass/fail), and another for hallucination detection via claim enumeration and assessment.
    \item A two-step process using separate prompts: one for well-formedness evaluation, and another for hallucination detection.
\end{enumerate}

To benchmark these judges, we constructed a dataset of 30 explanation samples, generated using smaller, older models and also manually-injected hallucinations, to ensure a high incidence of hallucinations and low-quality outputs. All samples were manually annotated using the scoring rubric.
The dataset included 17 acceptable samples, 9 samples with hallucinations, and 4 malformed samples. Figure~\ref{fig:example-hallucination} shows an example of a hallucinated statement. The codebase here is a tool for analyzing provisioned resources, not provisioning them. Meanwhile, Figure~\ref{fig:example-malformed} shows a malformed response in which the same sentence is repeated many times.
\begin{figure}[ht]
\begin{framed}
\scriptsize
\begin{quote}\texttt{
The code is a function called `validateVpcs` that takes a map of VPC details as input and returns an error if any of the VPCs have overlapping IP address spaces. The function iterates through each VPC and compares its address prefixes with the address prefixes of all other VPCs. If an overlap is found, an error is returned with a message indicating the overlapping VPCs. The purpose of the code is to ensure that VPCs do not have overlapping IP address spaces, which can cause connectivity issues. \textcolor{blue}{The code is part of a larger codebase that is used to generate infrastructure resources for IBM Cloud.} The GitHub pull requests referenced in the log history of the code are related to adding support for CIDR segments, modifying the model, converting IP and CIDR structs to IPBlock, and switching to the cobra CLI for parsing command line arguments.}
\end{quote}
\end{framed}
\caption{An example of a hallucination (in blue)}
\label{fig:example-hallucination}
\end{figure}

\begin{figure}[ht]
\begin{framed}
\scriptsize
\begin{quote}\texttt{
This commit disables envoy from using TLS session keys when connecting to the controlplane. It also disables the controlplane from using TLS session keys when connecting to the envoy. \textcolor{blue}{The commit also disables the envoy from using TLS session keys when connecting to the controlplane \dots}}
\end{quote}
\end{framed}
\caption{An example of a repetitive response}
\label{fig:example-malformed}
\end{figure}

\begin{figure}[th]
    \centering
    \includegraphics[width=1.0\linewidth]{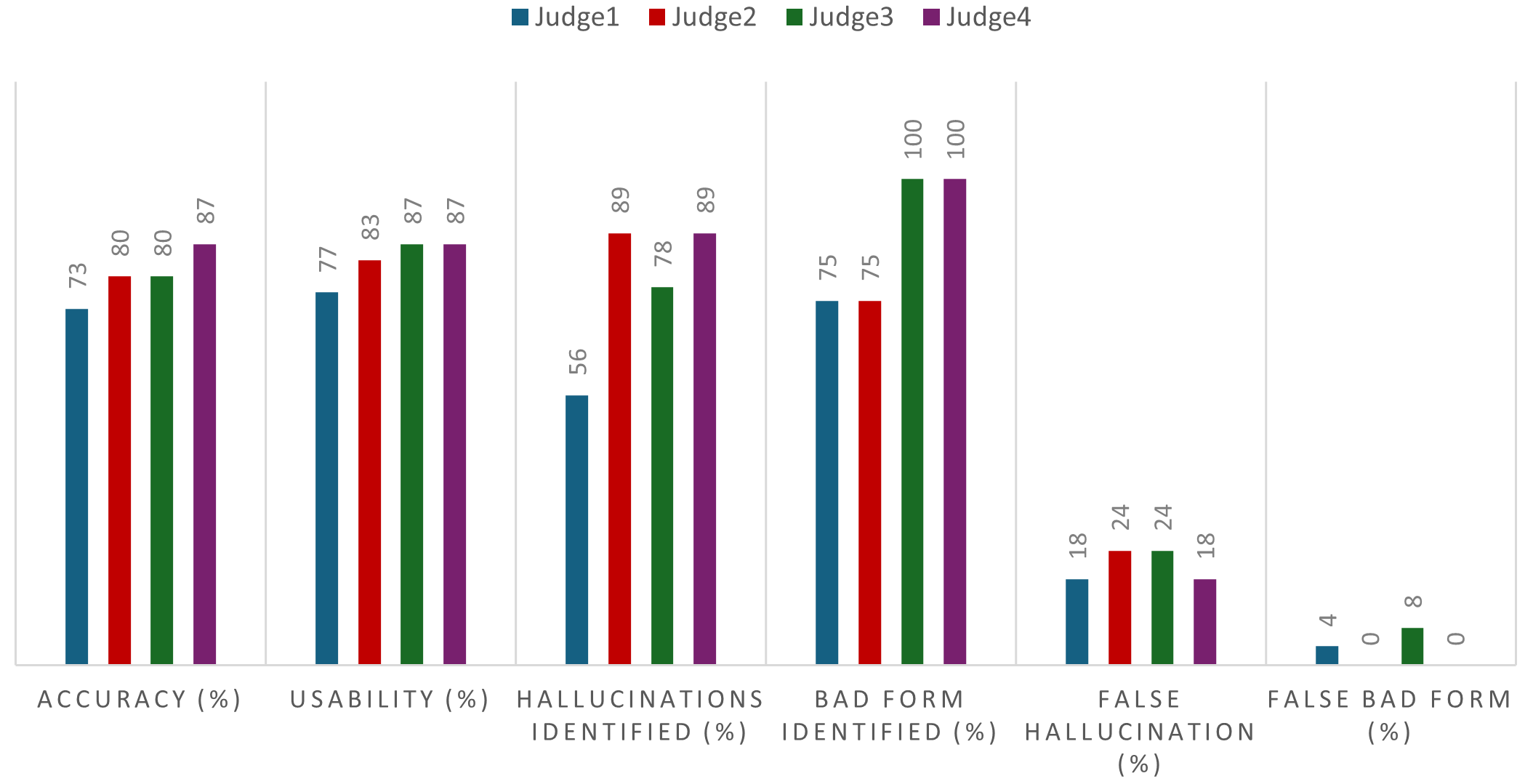}
    \caption{Evaluation results of different LLM judges.}
    \label{fig:judge-eval}
\end{figure}

Figure~\ref{fig:judge-eval} presents the evaluation results of four LLM-based judges (Judge1 through Judge4) across six metrics: Accuracy, Usability, Hallucinations Identified, Bad Form Identified, False Hallucination, and False Bad Form.

Accuracy measures the percentage of samples where the LLM judge's score matched the human annotation. For this metric, scores of 1 and 2 were considered equivalent, reflecting a shared recognition of hallucination regardless of severity. Usability evaluates whether the explanation is deemed fit for use: a response is considered usable if both the LLM and the human annotator agree that the explanation is either acceptable (score $=$ 0) or unacceptable (score $>$ 0).

Judge4, which employs a two-step evaluation process with separate prompts for well-formedness and hallucination detection, achieved the highest accuracy (87\%) and tied for the highest usability (87\%) with Judge3. Both Judge2 and Judge4 demonstrated superior performance in identifying hallucinations (89\%), while Judge3 and Judge4 were the only judges to consistently detect malformed explanations (100\%). Notably, Judge4 also maintained the lowest rate of false hallucination detection (18\%) and exhibited zero false positives in identifying bad form. These results support the conclusion that structured, multi-step evaluation—particularly when split across dedicated prompts—yields more reliable and interpretable judgments than single-pass approaches.

Our analysis yielded several key insights:

\begin{itemize}
    \item Explicit claim enumeration prior to hallucination assessment significantly improves evaluation accuracy.
    \item Prompt engineering plays a critical role in reducing false positives in hallucination detection.
    \item The two-step evaluation strategy (Judge4) consistently outperformed single-pass approaches in both precision and interpretability.
\end{itemize}

\subsection{Integration as an MCP Server}
The system can be deployed as an MCP server, exposing both the context extraction and explanation generation tools as modular services. This architecture enables seamless orchestration with other MCP components and supports scalable, on-demand usage.

The MCP server provides two primary endpoints:

\begin{itemize}
    \item \textbf{Context Extraction Tool:} Accepts a code snippet or function identifier and returns a structured bundle of GitHub-derived artifacts, including commit messages, PR descriptions, and linked issues.
    \item \textbf{LLM Explanation Tool:} Accepts both the code and the extracted context, and returns a high-level explanation of the code’s purpose within the application.
\end{itemize}

This service-oriented design offers several advantages:

\begin{itemize}
    \item \textbf{Reusability:} Each tool can be invoked independently or composed into larger workflows.
    \item \textbf{Configurability:} Supports runtime configuration of filters, token limits, and use-case-specific explanation styles.
    \item \textbf{Interoperability:} Compatible with other MCP tools for static analysis, transformation, and testing, enabling end-to-end modernization pipelines.
\end{itemize}

By exposing these tools as services, the MCP server facilitates integration into both interactive developer tools and automated agentic workflows, enhancing the accessibility and impact of GitHub-powered code understanding.

\section{User Experience Evaluation}
\label{evaluation}
Evaluating the purpose of a code snippet is inherently subjective and lacks a definitive ground truth. Human-authored explanations frequently incorporate information not present in the code or its associated GitHub context, and may emphasize different aspects depending on the author's perspective. Consequently, such explanations are unsuitable as gold-standard references for automated evaluation.

To assess the practical utility of our method, we conducted a user experience study involving six software repositories -- four open-source projects and two internal codebases. For each repository, we randomly selected between four and ten code snippets of varying sizes, ranging from a few lines to entire classes. For each snippet, we generated two explanations: one using only the code, and another augmented with GitHub-derived context. We then solicited feedback from project maintainers on the usefulness and accuracy of both explanations.

All participating developers identified at least one context-enhanced explanation that provided valuable insight beyond what was available from the code alone. Importantly, none of the real-world examples exhibited hallucinations. The only observed inaccuracies were cases where the explanation overemphasized details from pull request descriptions that were not central to the code’s purpose. Notably, several of the evaluated snippets originated from a large-scale project with hundreds of contextual artifacts, yet the LLM was consistently able to focus on the most relevant information.

As for runtime, almost no execution of our system lasted more than 20 seconds, typically less than 10 seconds. On the above-mentioned large-scale project, which carries more than 20 years of code history, one case was noted where runtime exceeded one minute. The code snippet to explain in this case had a history of 56 commits, out of which 38 were nontrivial. These commits were linked to 36 PRs, which in turn were linked to 95 issues. It is the fetching of all these PRs and issues that took most of the time, partly due to a less-performant GitHub Enterprise server.

These findings suggest that our approach is both scalable and effective in surfacing meaningful, context-aware explanations.

\section{Conclusions}
\label{conclusions}
In this work, we presented a system that enhances code understanding by leveraging contextual artifacts from GitHub and integrating them into a structured explanation workflow. Our approach combines a context extractor, an LLM-based explanation generator, and a novel LLM-as-a-Judge (LaaJ) validator to ensure the quality and reliability of generated insights. By prioritizing high-quality explanations and filtering out potentially misleading content, our system addresses a critical challenge in AI-assisted software comprehension.

Our system’s effectiveness is particularly evident in software maintenance scenarios, where understanding historical context and implementation rationale is essential. The GitHub-derived context often reveals valuable insights into application-level requirements and defect avoidance strategies.

The results of our user study validate the utility of our approach. Developers consistently found the insights to be a meaningful enhancement to standard code explanations, with no observed hallucinations. Importantly, the system maintains explanation integrity even when insights are unavailable, ensuring a non-disruptive user experience.

Overall, our work contributes a practical and reliable framework for integrating contextual code insights into developer workflows, paving the way for more informed and efficient software maintenance and evolution.

\section*{Acknowledgments}
We would like to thank Marah Ghoummaid for her valuable contributions in laying the foundations for the work on which this paper is based. Marah was an intern at IBM Research when she helped shape the direction of this work.

We would also like to thank Andreas Fried from IBM Software for his significant help in shaping the MCP-server integration.

\balance
\bibliographystyle{ieeetr}
\bibliography{references}

\end{document}